\documentclass[letterpaper, 10 pt, conference]{ieeeconf}
\IEEEoverridecommandlockouts
\usepackage[T1]{fontenc}
\usepackage{graphicx}
\usepackage{amssymb}
\usepackage{amsmath}
\usepackage{url}
\usepackage{makecell}
\let\labelindent\relax
\usepackage{enumitem}
\usepackage{booktabs}
\usepackage{microtype}

\usepackage{tikz}
\usetikzlibrary{arrows.meta,positioning,calc,fit,shapes.geometric}

\title{\LARGE \bf 
Progress-Based Fault Detection and Health-Aware Task Allocation \\ for Heterogeneous Multi-Robot Systems
}

\author{Jack Cline, Christian Macaranas, and Siavash Farzan 
\thanks{The authors are with the Electrical Engineering Department, California Polytechnic State University, San Luis Obispo, CA 93407, USA.}
\thanks{Corresponding author: {\tt\footnotesize sfarzan@calpoly.edu}}
}

\begin{document}

\maketitle
\thispagestyle{empty}
\pagestyle{empty}

\begin{abstract}
We present a progress-based fault detection module and its integration with dynamic task allocation for heterogeneous robot teams. The detector monitors a normalized task-completion signal with a lightweight Kalman filter (KF) and a normalized innovation squared (NIS) test, augmented with a low-rate stall gate, an uncertainty gate, and debounce logic. Health estimates influence the allocator via health-weighted costs and health-dependent masks; reallocation is event-triggered and regularized with an $\ell_1$ assignment-change penalty to limit reassignment churn while preserving feasibility through slack variables. The detector has constant per-robot update cost, and the allocation remains a convex quadratic program (QP). Experiments on a common team-task setup evaluate measurement-noise increases, velocity-slip biases, communication dropouts, and task abandonment. The results show timely detection in the noise and bias cases, maintained task completion with limited reassignment, and the expected observability delays under communication dropouts.
\end{abstract}

\section{Introduction}
Heterogeneous multi-robot teams are increasingly deployed for logistics, inspection, and field operations where reliability and continuity of service are mandatory~\cite{Korsah2013}. In these settings, robots experience intermittent sensing noise, actuation limits, and communication dropouts that degrade task progress and can cascade into missed deadlines and idle resources. Dynamic task allocation can mitigate these effects only if the system can quickly identify degrading agents and adjust assignments without destabilizing the overall schedule~\cite{Parker1998}. This work targets that gap by designing a lightweight, progress-based fault detection module that operates online with constant per-robot cost, and by integrating its health estimates into a dynamic allocation program to reassign work with minimal disruption, bounded assignment churn, and preserved feasibility.
We make two technical contributions and an empirical study:
\begin{list}{}{\leftmargin=0em \itemindent=6pt}
    \item[i.] \textit{Progress-based fault detection:} We estimate normalized task progress and its rate with a lightweight two-state discrete Kalman filter, and detect anomalies using a normalized innovation squared $\chi^2_1$ test. Robustness is improved with a low progress-rate gate, an uncertainty gate, and debounce/hysteresis. A windowed NIS statistic increases sensitivity to persistent deviations while controlling false alarms.
    \item[ii.] \textit{Health-aware integration with dynamic allocation:} We map health estimates into allocation signals through health-weighted costs and health-dependent gating. Reallocation is event-triggered on health changes and regularized with an $\ell_1$ assignment-change penalty to limit reassignment churn while preserving feasibility via slack.
    \item[iii.] \textit{Empirical validation:} We evaluate representative fault families (measurement-noise increases, velocity-slip bias, communication dropout, and task abandonment) on a common heterogeneous team-task setup. We report detection delay, task completion under reallocation, healthy-stream innovation statistics, and fault-magnitude sweep results summarized via ROC curves and accuracy tables.
\end{list}

\section{Related Work}

\subsection{Heterogeneous Task Allocation}
Dynamic task allocation for heterogeneous teams is often posed as a convex program, with capability constraints captured by specialization matrices and task priorities encoded as weights~\cite{6729278}. Feasibility under disturbances and modeling error is maintained through bounded slack and soft penalties that relax hard constraints when needed~\cite{TA_review}. Quadratic formulations enable efficient real-time re-solving as tasks and robot states change, which is essential for online operation~\cite{8795895}. Recent data-driven schemes adjust weights or feasibility sets on the fly to reflect evolving capabilities and environment conditions~\cite{9560857}.

\subsection{Fault Detection and Diagnosis for Robot Teams}
Innovation and residual tests from state estimation are a common basis for anomaly detection in robotics; Kalman filtering provides residuals with known statistics for principled decisions~\cite{thrun2005probabilistic}. The normalized-innovation-squared test offers a $\chi^2$ decision rule suitable for online operation and is widely used in practice~\cite{grewal2014kalman}. Beyond residual tests, generalized likelihood ratio and CUSUM-style change detectors capture abrupt faults and slow drifts in team settings~\cite{s21196536}. Command-aware filtering that incorporates nominal inputs or expected progress rates improves separability between modeling error and true faults~\cite{FD_and_D}. Recent surveys emphasize lightweight, online detectors that tolerate missing data and noise while interfacing cleanly with coordination layers~\cite{faults_review}. A broader review of multi-robot fault diagnosis and fault-tolerant control discusses sensitivity-false-alarm trade-offs under communication constraints~\cite{quamar2024reviewfaultdiagnosisfaulttolerant}.

\subsection{Fault-Tolerant Allocation}
Once faults are detected, allocation can react through soft demotion, constraint gating, or temporary removal of affected agents, to preserve feasibility and task coverage under degraded capability~\cite{ASMO}. Optimization-based approaches implement these responses by shaping costs or inserting assignment masks so that faulty robots are deprioritized without making the problem infeasible~\cite{10127604}. 

Beyond optimization, fault tolerance has been expressed with behavior-tree abstractions that switch or degrade actions as health decreases to enable modular recovery at the task layer~\cite{colledanchise2015adaptivefaulttolerantexecution}. Resilient coordination primitives further maintain performance under agent loss or communication faults by adapting selection and consensus mechanisms~\cite{bossens2022resilient}. 
To curb oscillations during repeated reallocations, several methods penalize reassignment changes or enforce short cooldowns, which reduces churn while sustaining mission progress~\cite{10539334}.

\section{Preliminaries and Problem Formulation}\label{sec:preliminaries}

\subsection{Notation and System Model}
Let $\mathcal{R} = \{1,\dots,N\}$ be the set of robots and $\mathcal{T} = \{1,\dots,M\}$ the set of tasks. Each robot $i$ has state $\mathbf{x}_i$, control input $\mathbf{u}_i$, specialization matrix $S_i$ that encodes capability for each task, priority weights $\alpha_{i,m}\in[0,1]$, and slack variables $\delta_{i,m}\ge 0$ that preserve feasibility. The sampling time is $\Delta t$. For each robot-task pair we define a normalized task progress signal $r_{i,m}(t)\in[0,1]$ and a noisy measurement $y_{i,m}(t)$ used by the detector and estimator.

\subsection{Dynamic Heterogeneous Task Allocation}
We summarize the dynamic allocation problem~\cite{8795895} used in our framework and establish the notation needed in later sections. At each allocation update, the optimizer, formulated as a quadratic program (QP), computes control inputs and assignment variables that balance control effort, task priorities, and feasibility slack while respecting specialization and safety constraints:
\begin{align}
    \min_{\{\mathbf{u}_i,\boldsymbol{\delta}_i,\boldsymbol{\alpha}_i\}} \quad 
    & \sum_{i=1}^N \Big( \|\mathbf{u}_i\|_2^2 + c^\top \boldsymbol{\delta}_i \Big) 
    + \sum_{i=1}^N \sum_{m=1}^M w_{i,m}\, \alpha_{i,m} \label{eq:alloc_obj}\\
    \text{s.t.} \quad 
    & \frac{\partial h_{i,m}}{\partial \mathbf{x}_i}\, \mathbf{u}_i \ge -\kappa\, h_{i,m} - \rho\, \delta_{i,m}, \quad \forall i,m \label{eq:cbf}\\
    & \alpha_{i,m} \le S_i(m), \quad \forall i,m \label{eq:spec}\\
    & \mathbf{1}^\top \boldsymbol{\alpha}_i \le 1, \quad \boldsymbol{\alpha}_i \in [0,1]^M, \quad \forall i \label{eq:assign}\\
    & \|\boldsymbol{\delta}_i\|_\infty \le \delta_{\max}, \quad \boldsymbol{\delta}_i \ge 0, \quad \forall i. \label{eq:slack}
\end{align}
Here $h_{i,m}(\mathbf{x}_i)$ is a task-dependent barrier or progress function whose positive values correspond to safe or desired operation. The slack variable $\delta_{i,m}$ softens \eqref{eq:cbf}: when $\delta_{i,m}>0$, temporary violation of the nominal inequality is permitted, but it is penalized in the objective through $c^\top \boldsymbol{\delta}_i$. This is the mechanism that preserves feasibility when strict satisfaction of all task constraints is impossible. The specific form of $h_{i,m}$ is problem-dependent. For spatial goal-reaching tasks, $h_{i,m}$ can be chosen from a smooth distance-to-goal or safety-buffer function; for other tasks, any differentiable surrogate whose increase corresponds to task progress can be used.

The QP is re-solved online at each allocation epoch or when health labels change. The formulation is myopic rather than horizon-based: it has no explicit prediction horizon, and time enters through the repeated re-solving of the QP as the states, measurements, weights, and masks evolve.

\subsection{Fault Model and Objectives}
We consider operational degradations that manifest as changes in task progress production or measurement, including biased sensing, progress stalls, actuation limits, and intermittent communication. Rather than modeling low-level physics of each fault, we capture their effect through the statistics of $r_{i,m}(t)$ and $y_{i,m}(t)$ used by the estimator and detector. The objectives are fast detection with controlled false-alarm rate, stable health-to-allocation mapping that preserves feasibility through \eqref{eq:slack}, and minimal assignment churn while maintaining task coverage and safety~\cite{FD_and_D,faults_review,quamar2024reviewfaultdiagnosisfaulttolerant}.\looseness=-1

Our goal is not to replace platform-specific fault detection at the physics level, but to complement it with a task-level monitor that can be shared across heterogeneous robots. A task-progress signal can be defined even when robots have different state dimensions, controllers, and sensing modalities, whereas a physics-level residual is usually tied to a specific vehicle model. Monitoring progress therefore provides a common interface between detection and allocation: different low-level faults are mapped to the same operational question, namely whether the robot is producing the task completion expected by the coordinator.

\section{Progress-Based Fault Detection}\label{sec:progress_detection}

\subsection{Progress Signal Construction}
Conceptually, the progress signal measures useful task completion rather than low-level motion. It answers the coordinator's question: how much of the assigned task has this robot completed so far? This task-level abstraction is useful for heterogeneous teams because it can be defined even when robots have different dynamics, controllers, or sensing suites.
We define a normalized task progress signal $r_{i,m}(t)\in[0,1]$ for each robot-task pair, sampled at interval $\Delta t$. The construction depends on task type:
\begin{itemize}[leftmargin=*]
    \item \textit{Spatial tasks} (coverage, patrolling, path following): $r(t)$ is the traversed distance or covered area normalized by the planned total, with optional geometric clipping near endpoints to reduce quantization effects. 
    \item \textit{Workload tasks} (queues, pick-and-place, inspection lists): $r(t)$ is the completed fraction $\frac{\text{items done}}{\text{items assigned}}$ or a smoothed count of \emph{completion events}, where a completion event means the successful completion of one discrete work unit such as serving one queued request, picking one item, or inspecting one listed location.
    \item \textit{Composite tasks}: $r(t)$ is a soft aggregation of subtask progresses, for example a weighted sum subject to clamping $r(t)\leftarrow\min\{1,\max\{0,\,\sum_k w_k r_k(t)\}\}$, or a smooth-min when the overall task is bottlenecked by the slowest subtask.
\end{itemize}
These categories are representative rather than exhaustive. More generally, any task that admits a monotone or piecewise-monotone completion surrogate in $[0,1]$ can be monitored by the proposed detector.

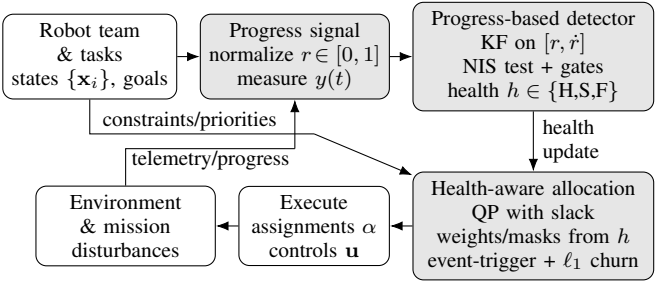
\begin{figure}[t]
\centering
\begin{tikzpicture}[
    font=\footnotesize,
    >=Latex,
    node distance=8mm and 3mm,
    block/.style={draw, rounded corners, align=center, minimum width=20mm, minimum height=9mm},
    blockw/.style={draw, rounded corners, align=center, minimum width=32mm, minimum height=9mm}
]

\node[block] (team) {Robot team\\\& tasks\\{\footnotesize states $\{\mathbf{x}_i\}$, goals}};
\node[block] (progress) [right=of team, fill=black!10] {Progress signal\\{\footnotesize normalize $r\!\in[0,1]$}\\{\footnotesize measure $y(t)$}};
\node[blockw] (detector) [right=of progress, fill=black!10] {Progress-based detector\\{\footnotesize KF on $[r,\dot r]$}\\{\footnotesize NIS test + gates}\\{\footnotesize health $h\in\{$H,S,F$\}$}};

\node[blockw] (allocator) [below=of detector, fill=black!10] {Health-aware allocation\\{\footnotesize QP with slack}\\{\footnotesize weights/masks from $h$}\\{\footnotesize event-trigger + $\ell_1$ churn}};
\node[block] (execute) [left=of allocator] {Execute\\{\footnotesize assignments $\alpha$}\\{\footnotesize controls $\mathbf{u}$}};
\node[block] (env) [left=of execute, text width=2.15cm] {Environment\\\& mission \\ disturbances}; 

\draw[->] (team) -- (progress);
\draw[->] (progress) -- (detector);
\draw[->] (detector) -- node[pos=0.5, right, align=left] {\footnotesize health \\ update} (allocator);
\draw[->] (allocator) -- (execute);
\draw[->] (execute) -- (env);

\node [coordinate, below of=progress, node distance=12.25mm] (coord1) {};
\draw (env.north) |- node[pos=0.35, left, align=left,xshift=2.3cm] {\footnotesize telemetry/progress} (coord1);
\draw[->] (coord1) -- (progress.south);

\node [coordinate, above of=execute, node distance=12.25mm] (coord2) {};
\draw (team.south) |- (coord2);
\draw[->] (coord2) -- node[pos=0.15, above, align=left,xshift=-1.85cm] {\footnotesize constraints/priorities} (allocator);
\end{tikzpicture}
\vspace{-10pt}
\caption{Overview of the proposed methodology: progress is monitored to infer robot health, which conditions allocation decisions; execution affects the environment and produces new progress measurements.}
\label{fig:method_overview_detailed}
\end{figure}

\subsection{Estimator}
We employ a lightweight discrete Kalman filter on the state $x_t=\begin{bmatrix} r_t & \dot r_t \end{bmatrix}^\top$ with two interchangeable models:
\begin{enumerate}[leftmargin=*]
    \item \textit{Command-unaware constant-velocity (CV)}:
    \begin{align}
        x_{t+1} &= 
        \underbrace{\begin{bmatrix}1 & \Delta t\\[2pt] 0 & 1\end{bmatrix}}_{A_{\text{cv}}}
        x_t + w_t, \qquad w_t\sim\mathcal{N}(0,Q), \label{eq:cv_dynamics}\\
        y_t &= 
        \underbrace{\begin{bmatrix}1 & 0\end{bmatrix}}_{C} x_t + v_t, \qquad v_t\sim\mathcal{N}(0,R). \label{eq:meas}
    \end{align}
    \item \textit{Command-aware rate-tracking (RT)} when a nominal progress rate $u_t$ is available from the allocator/supervisor:
    \begin{equation}
        x_{t+1} = 
        \underbrace{\begin{bmatrix}1 & 0\\[2pt] 0 & 0\end{bmatrix}}_{A_{\text{rt}}}
        x_t + 
        \underbrace{\begin{bmatrix}\Delta t\\[2pt] 1\end{bmatrix}}_{B}
        u_t + w_t, \quad
        y_t = C x_t + v_t.  \label{eq:rt_dynamics}
    \end{equation}
\end{enumerate}
The CV model assumes locally constant $\dot r$, while the RT model treats $u_t$ as the nominal rate and absorbs modeling errors into $w_t$. Both filters have $O(1)$ per-robot update cost and can be tuned by $Q$ and $R$ based on observed noise and variability. Standard predict-update recursions provide the innovation $\tilde y_t$ and the a priori covariance $P_t^{-}$ needed by the detector~\cite{NDKF2025}.\looseness=-1

In the KF, $R$ models uncertainty in the measured progress signal and $Q$ captures unmodeled variability in task execution. In practice, we initialize $R$ from sensor-noise propagation into the normalized progress coordinate and then tune $Q$ so that healthy-stream innovations are approximately white and the mean NIS on healthy runs is close to one. This calibration step is important because the progress dynamics are task-dependent and need not be known exactly.

\subsection{NIS Decision Rule with Operational Gates}
We use the innovation (the one-step-ahead measurement prediction error) rather than a generic residual. Specifically,\looseness=-1
\begin{align}
    \tilde y_t &= y_t - C \hat x_t^{-}, \\
    S_t &= C P_t^{-} C^\top + R, \\
    d_t &= \tilde y_t^\top S_t^{-1} \tilde y_t .
\end{align}
Here $\tilde y_t$ is the innovation, $S_t$ is its covariance, and $d_t$ is the normalized innovation squared (NIS). Throughout the paper, $\chi^2_1$ denotes the chi-square distribution with one degree of freedom. Under the standard linear-Gaussian KF assumptions and for a scalar measurement, $d_t$ is approximately $\chi^2_1$-distributed when the filter is well calibrated. We therefore choose a significance level $\alpha\in(0,1)$ and set the decision threshold as
\( \tau_\alpha = F^{-1}_{\chi^2_1}(1-\alpha), \)
where $F^{-1}_{\chi^2_1}(\cdot)$ is the inverse cumulative distribution function of $\chi^2_1$.

To improve robustness in online operation, we augment the NIS test with three operational gates:
\begin{itemize}[leftmargin=*]
    \item \textit{Low progress-rate gate}: if $|\widehat{\dot r}_t|<\beta$ for $L$ consecutive samples, where $\beta>0$ is a stall threshold, the stream is flagged as stalled even if $d_t\le \tau_\alpha$.
    \item \textit{Uncertainty gate}: if $\mathrm{trace}(P_t)>\gamma$, where $\gamma>0$ is a user-selected covariance threshold, suspend new fault declarations and mark the stream as temporarily uninformative.
    \item \textit{Debounce/hysteresis}: let $z_t\in\{0,1\}$ denote whether either the NIS test or the low-rate gate is active at time $t$. If $z_t=1$ for $K_s$ consecutive samples, the label changes from healthy to \emph{suspect}; if the trigger persists for $K_f>K_s$ consecutive samples, the label changes from suspect to \emph{fault}. The label returns toward healthy only after $K_{\text{out}}$ consecutive non-trigger samples.
\end{itemize}

The detector also outputs a confidence score
\begin{equation}
    c_t=\min\!\left\{1,\max\!\left\{0,\frac{d_t/\tau_\alpha-1}{\eta}\right\}\right\},
\end{equation}
where $\eta>0$ sets the slope of the confidence map.

\subsection{Properties}
Under approximately independent innovations, requiring $K_s$ consecutive triggers reduces the effective per-step false-alarm rate from $\alpha$ to roughly $\alpha^{K_s}$, at the cost of at least $K_s-1$ additional samples of detection delay after threshold crossing. Increasing the window size $w$ or the stall dwell length $L$ produces a similar trade-off: larger values reduce false alarms but increase latency. We use this relationship in tuning to balance sensitivity against nuisance alarms.

\subsection{Communication Dropout Analysis}
When progress measurements are unavailable for $D$ consecutive steps, the filter runs in prediction mode only and the covariance evolves as
\begin{equation}
    P_{k+D|k} = A^D P_{k|k} A^{D\top} + \sum_{j=0}^{D-1} A^j Q A^{j\top}.
\end{equation}
For the constant-velocity model $A=A_{\mathrm{cv}}$, we have
\begin{equation}
    A^D = \begin{bmatrix} 1 & D\Delta t \\ 0 & 1 \end{bmatrix},
\end{equation}
so $\mathrm{trace}(P_{k+D|k})$ grows with the outage length $D$. We therefore define the maximum informative outage length as
\begin{equation}
    D_{\max} = \max \left\{ D : \mathrm{trace}(P_{k+D|k}) \le \gamma \right\}.
\end{equation}
Once $D>D_{\max}$, the uncertainty gate suppresses new fault declarations and the system degrades conservatively to an uncertain monitoring mode. Thus, the limiting mechanism under communication delay is covariance growth rather than instability of the allocator. This observation is consistent with the communication-dropout experiments, where outages lasting tens of samples delayed detection substantially.

\section{Integration with Dynamic Allocation}\label{sec:integration}

\subsection{Health-Weighted Costs}
Let $h_i(t)\in\{\text{healthy},\text{suspect},\text{fault}\}$ denote the health label of robot $i$. In Fig.~\ref{fig:method_overview_detailed}, these labels are abbreviated as $H$, $S$, and $F$, respectively.
The key design choice is to \emph{demote} degraded robots before \emph{excluding} them entirely. A suspect label indicates evidence of degraded performance but not yet enough certainty for irreversible removal; immediate exclusion at this stage can create unnecessary coverage loss or even infeasibility. We therefore first modify the allocation objective so that healthier substitutes are preferred whenever they exist:\looseness=-1
\begin{align}
    w_{i,m}(t) &= w^{0}_{i,m} + \kappa_s\,\mathbf{1}\{h_i(t)=\text{suspect}\} \nonumber \\
    &\qquad\quad + M_f\,\mathbf{1}\{h_i(t)=\text{fault}\}, \label{eq:weight_schedule}\\
    c_i(t) &= c_i^{0}\,\big(1 + \rho_s\,\mathbf{1}\{h_i(t)=\text{suspect}\}\big). \label{eq:slack_schedule}
\end{align}
Here $w^{0}_{i,m}$ and $c_i^{0}$ are nominal weights, $\kappa_s>0$ is a moderate demotion penalty for suspect robots, and $M_f\gg 0$ is a large penalty for confirmed faults. This layered schedule preserves convexity while supporting graceful degradation: suspect robots are used only when needed, whereas confirmed faults are either strongly discouraged or excluded altogether via masks. This is preferable to removing all degraded agents immediately, because it lets the optimizer trade off health, coverage, and feasibility in a principled way.

\subsection{Constraint Gating and Masks}
For critical tasks or safety conditions, we gate assignments using a health-dependent mask that is treated as a parameter during optimization:
\begin{gather}
    \alpha_{i,m}(t) \le S_i(m)\, g_i(t), \label{eq:mask}\\
    \text{with} \quad g_i(t) = \begin{cases}
    1, & h_i(t)\in\{\text{healthy},\text{suspect}\},\\
    0, & h_i(t)=\text{fault}, \nonumber
    \end{cases} 
\end{gather}
so that confirmed faults cannot be bound to gated tasks, while feasibility is maintained via \eqref{eq:slack}. This preserves the convex, QP structure because $g_i(t)$ is fixed at solve time. Where hard exclusion is not required, a large penalty $M_f$ in \eqref{eq:weight_schedule} implements a soft mask. Such gating and soft-hard isolation mechanisms are standard in fault-tolerant behavior composition and resilient coordination~\cite{ASMO,colledanchise2015adaptivefaulttolerantexecution}.

\subsection{Event-Triggered Reallocation with Churn Control}
We re-solve the allocation when any $h_i(t)$ changes or at a slow periodic rate to refresh warm starts. Let
\begin{equation}
    J(\mathbf{u},\boldsymbol{\delta},\boldsymbol{\alpha})
    =
    \sum_{i=1}^N \Big(\|\mathbf{u}_i\|_2^2 + c_i(t)^\top \boldsymbol{\delta}_i \Big)
    + \sum_{i=1}^N \sum_{m=1}^M w_{i,m}(t)\,\alpha_{i,m}
\end{equation}
denote the baseline allocation objective. We define \emph{assignment churn} as the total change in robot-task assignments between two consecutive solves. If $\mathbf{a}_t$ stacks $\{\alpha_{i,m}(t)\}_{i,m}$ and $\bar{\mathbf{a}}_{t-1}$ is the previous solution, we use
\begin{equation}
    J_{\text{total}} = J + \lambda \, \|\mathbf{a}_t - \bar{\mathbf{a}}_{t-1}\|_1, \label{eq:churn}
\end{equation}
where $\lambda\ge 0$ tunes the trade-off between responsiveness and stability. The $\ell_1$ penalty promotes sparse assignment updates, meaning that the optimizer tends to change only the few assignments needed to accommodate a detected fault. Compared with an $\ell_2$ penalty, this produces more interpretable reallocations and reduces oscillatory, small-amplitude changes spread across many assignments.

\subsection{Algorithmic Pipeline}
We run the following lightweight detector-and-allocator loop at each sampling step, which links progress sensing, fault detection, and health-aware allocation:
\begin{enumerate}[leftmargin=*]
    \item Read progress measurements $y_t$ and run the KF update to obtain $(\hat x_t,P_t)$.
    \item Compute the NIS statistic and apply gates to update $h_i(t)$ and a confidence score.
    \item Form weights $w_{i,m}(t)$ and penalties $c_i(t)$ via \eqref{eq:weight_schedule}-\eqref{eq:slack_schedule}, and masks $g_i(t)$ via \eqref{eq:mask}.
    \item Warm start the QP with $\bar{\mathbf{a}}_{t-1}$ and solve the allocation with the churn cost \eqref{eq:churn}.
    \item Execute assignments, store $\bar{\mathbf{a}}_{t}\hspace{-1.5pt}\leftarrow\hspace{-2pt} \mathbf{a}_t$, and update cooldowns.
\end{enumerate}

\subsection{Computational Considerations}
The detector adds constant per-robot overhead and does not affect convexity. The allocation remains a QP with health-dependent parameters as in \eqref{eq:alloc_obj}-\eqref{eq:slack}; its complexity scales with the number of robots, tasks, and control variables as in standard dynamic heterogeneous allocation~\cite{8795895,MRTA_review}. In practice, warm starts and sparse updates in \eqref{eq:churn} reduce iteration counts to support real-time replanning in multi-robot deployments.\looseness=-1

\section{Simulation Results and Discussions}

\subsection{Setup}\label{subsec:setup}
\begin{figure}[!hb]
    \centering
    \includegraphics[trim={10pt 15pt 20pt 30pt},clip,width=0.8\linewidth]{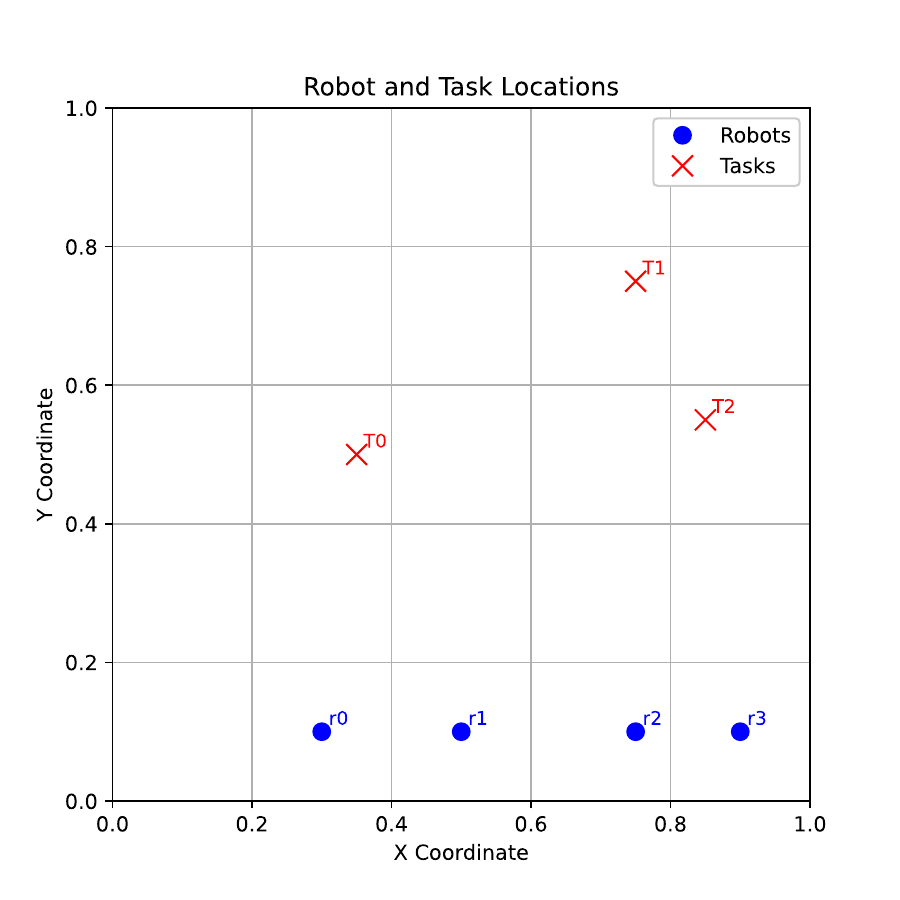}
    \caption{Initial robot and task positions (in meters) used across scenarios. Tasks are placed at $(0.35,0.50)$, $(0.75,0.75)$, and $(0.85,0.55)$. Robots are initially placed at $(0.3, 0.1)$, $(0.5, 0.1)$, $(0.75, 0.1)$, and $(0.9, 0.1)$.}
    \label{fig:init_positions}
\end{figure}

We evaluate the detector-allocator pipeline in a spatial, three-task scenario with four heterogeneous robots. The simulator was implemented in Python using the \textit{quadprog} library for the QP and a fixed sampling time of 0.1 s. Robots start near the lower boundary of the workspace and must reach fixed task goals (Fig.~\ref{fig:init_positions}). The chosen robot-task geometry creates distinct nominal distances and progress profiles across robot-task pairs, which makes it possible to study both detection behavior and reassignment decisions in a controlled setting.

Table~\ref{tab:robot_task_positions_acc} lists normalized distances from each robot to each task at $t=0$, which shape progress observability and the covariance scaling described below.

Unless stated otherwise, the specialization matrix is
\[
    S_i=\begin{bmatrix}
    1 & 0 & 0\\[-2pt]
    0 & 1 & 0\\[-2pt]
    0 & 0 & 1\\[-2pt]
    0 & 0 & 1
    \end{bmatrix},
\]
which assigns Robots 0-2 uniquely to Tasks 0-2, with Robot 3 additionally specialized for Task 2 to provide redundancy during reallocation. Scenario-specific variants are noted where used.\looseness=-1

\begin{table}[!t]
\centering
\caption{Three-task setup: normalized robot-task distances at $t=0$.}
\label{tab:robot_task_positions_acc}
\begin{tabular}{lcccc}
    \toprule
    \textbf{Task $\backslash$ Robot} & \textbf{Robot 0} & \textbf{Robot 1} & \textbf{Robot 2} & \textbf{Robot 3} \\
    \midrule
    \textbf{Task 0 (0.35, 0.50)} & 0.403 & 0.412 & 0.513 & 0.557 \\
    \textbf{Task 1 (0.75, 0.75)} & 0.794 & 0.728 & 0.650 & 0.671 \\
    \textbf{Task 2 (0.85, 0.55)} & 0.666 & 0.550 & 0.458 & 0.452 \\
    \bottomrule
\end{tabular}
\end{table}

We adopt the progress-state models in Section~\ref{sec:progress_detection} and tune estimator covariances using distance-dependent scaling so that innovation magnitudes are comparable across tasks of different spatial scales. With position sensor noise $\sigma_{xy}=0.007$, normalized measurement variance for progress is
\begin{equation}
    R = \frac{\sigma_{xy}^2}{L^2}=\frac{4.9\times 10^{-5}}{L^2},
\end{equation}
where $L$ is the robot’s current distance to its assigned task goal. The baseline process covariance is
\begin{equation}
    Q_0=\frac{1}{L^2}\begin{bmatrix}
    1.0\times 10^{-6} & 0\\
    0 & 5.0\times 10^{-5}
    \end{bmatrix},
\end{equation}
with adaptive scaling during operation to match observed variability. The NIS uses a window of $w=15$ samples for stability, and debounce is applied as in Section~\ref{sec:progress_detection}.

We test four representative fault families:
(i) \emph{Measurement noise increase}: progress measurement noise rises at step 20 and remains elevated.
(ii) \emph{Velocity slip bias}: a systematic progress-rate bias is injected on a selected robot at step 20.
(iii) \emph{Communication dropout}: progress measurements are dropped starting at step 10 to reveal the covariance growth mechanism.
(iv) \emph{Task abandonment}: a robot is immobilized for a contiguous window to emulate actuator failure, which exercises the rate gate and debounce. 
Across scenarios, dynamic allocation is solved as a QP with health-weighted costs and constraint gating (Sections~\ref{sec:integration} and~\ref{sec:preliminaries}), using a standard dense QP solver. The program structure follows heterogeneous, convex multi-robot task-allocation formulations~\cite{MRTA_review}.

\subsection{Metrics, Baselines, and Ablations}\label{subsec:metrics}

We report detector, allocator, and system-level metrics:
\begin{itemize}[leftmargin=*]
    \item \textit{Detection delay} (steps from injection to declaration) and \textit{false-alarm rate} per robot-task stream, using NIS criteria.
    \item \textit{Innovation sanity} via mean and std. dev. of NIS on healthy streams, which targets unit mean under correct modeling.
    \item \textit{Task completion} rate and time, and \textit{slack usage} as a feasibility surrogate for degraded operation.
    \item \textit{Assignment churn} $\|\mathbf a_t-\mathbf a_{t-1}\|_1$ and reallocation frequency to reflect stability of the integration scheme.
    \item \textit{Computational load}: detector update time per robot and QP solve time per step.
    \item \textit{ROC/AUC} and accuracy over fault-magnitude sweeps for noise and bias.
\end{itemize}

We compare against:
\begin{itemize}[leftmargin=*]
    \item \textit{No-FD}: allocation without health inputs.
    \item \textit{Naive thresholds on progress or rate}: declare fault if $r_t<\theta_r$ or $|\dot r_t|<\theta_{\dot r}$ for $L$ steps.
    \item \textit{Oracle FD}: perfect labels injected to the allocator to isolate the integration effect.
\end{itemize}
Ablations isolate the role of each component:
\begin{itemize}[leftmargin=*]
    \item Remove the \textit{rate gate} (stall detection) or the \textit{uncertainty gate} to keep the NIS core.
    \item \textit{Weights only} versus \textit{hard masks} for confirmed faults in the allocator.
    \item Disable the \textit{churn regularizer} to quantify its effect on reassignment stability.
\end{itemize}

\subsection{Results}\label{subsec:results}
\begin{figure}[t]
    \centering
    \includegraphics[width=\linewidth]{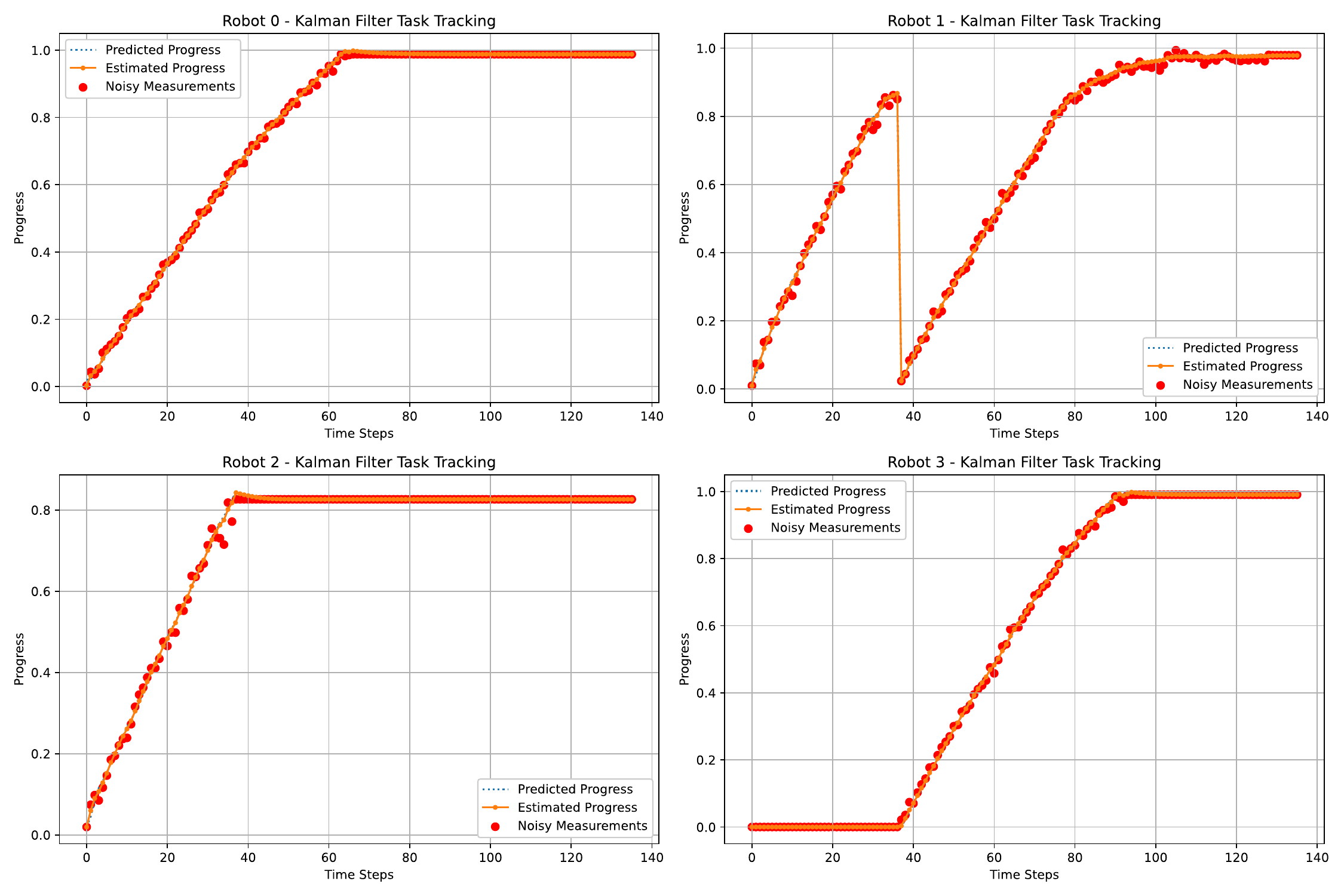} \\[-1pt]
    \footnotesize{(a) Normalized task progress $r(t)$} \\[-1pt]
    \includegraphics[width=\linewidth]{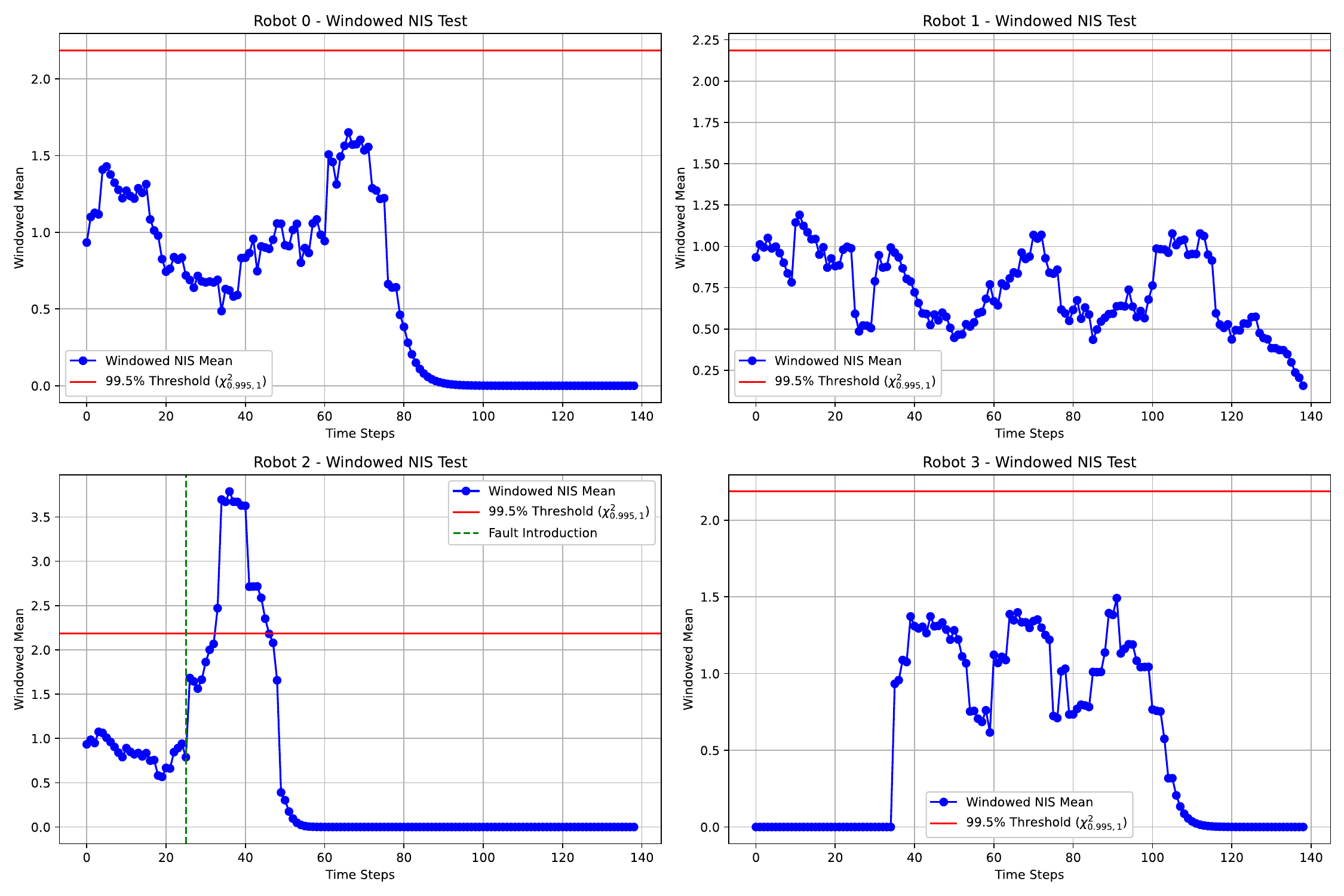} \\[-1pt]
    \footnotesize{(b) Windowed NIS statistic}
    \caption{Representative measurement-noise fault experiment for robots $i\in\{0,1,2,3\}$. After the measurement-noise increase is injected on Robot 2, the NIS crosses the detection threshold, triggering the integration layer to reallocate the affected task while maintaining limited reassignment churn.}
    \label{fig:rep_run_noise}
\end{figure}

We evaluate the detector-allocator pipeline on four representative fault families using the common team and task setup from Section~\ref{subsec:setup}: measurement-noise increase, velocity slip bias, communication dropout, and task abandonment.
Fig.~\ref{fig:rep_run_noise}(a) shows the normalized progress trace for the representative measurement-noise run, while Fig.~\ref{fig:rep_run_noise}(b) shows the corresponding windowed NIS statistic with detection markers.
After the noise increase is injected, the progress signal becomes more erratic and the NIS crosses the threshold, which triggers a fault declaration and subsequent health-aware reallocation. The NIS remains near its healthy baseline before the fault, crosses the threshold at the declared detection time, and stays elevated until the health-aware reallocation takes effect. This behavior is consistent with the intended role of the window and debounce parameters: they suppress isolated threshold crossings while still allowing persistent faults to trigger reallocation.

This test employs the following specialization matrix:
\[
S_i=\bigl[\,[0,1,0];\ [1,0,1];\ [0,0,1];\ [1,0,0]\,\bigr].
\]

Table~\ref{table:noise_summary} quantifies this scenario. The injected measurement-noise fault on Robot 2 at step 25 is detected at step 36 (11-step delay), after which Robot 2 no longer receives assignments. Prioritization of specialization constraints is demonstrated by dynamically reallocating Robot 1 to Task 2 and Robot 3 to Task 0 upon Robot 2 failing, illustrating the adaptability of the framework.  The NIS statistics for healthy robots remain close to unity (e.g., 0.83-0.96 average), which indicate well-tuned innovation magnitudes in the absence of faults.

Table~\ref{table:bias_summary_acc} reports the velocity slip bias experiment. A systematic progress-rate bias applied to Robot 1 at step 20 is detected at step 28. Following detection, assignments are updated so that Robot 3 covers Task 1, while other robots maintain their tasks. Again, healthy-stream NIS averages are near one (0.93-1.10), which support the validity of the residual-based test when the model is accurate.

The velocity-slip test uses this specialization matrix:
\[
S_i=\bigl[\,[0,0,1];\ [0,1,0];\ [1,0,0];\ [0,1,0]\,\bigr].
\]

\begin{figure}[!t]
    \vspace{-12pt}
    \centering
    \includegraphics[trim={20pt 15pt 50pt 20pt},clip,width=0.492\linewidth]{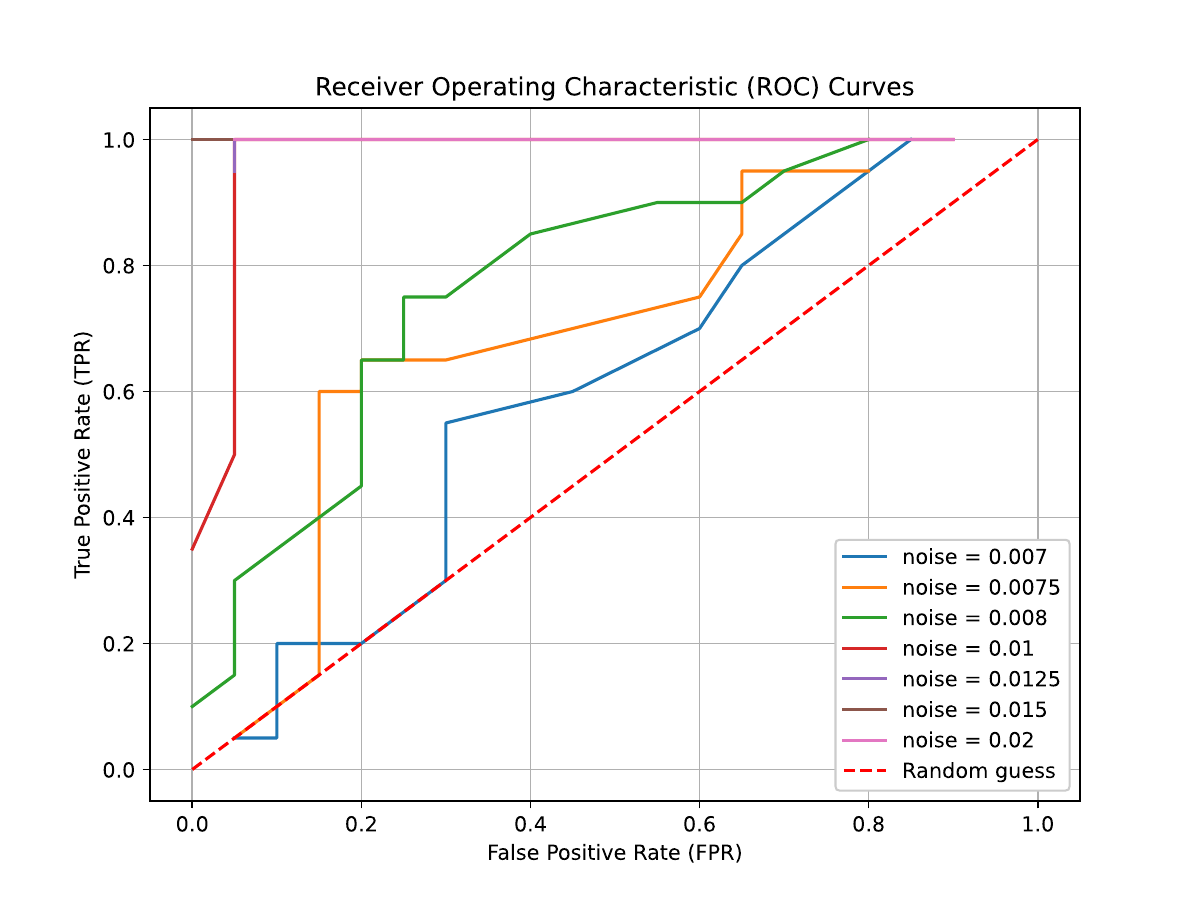}
    \includegraphics[trim={20pt 15pt 50pt 20pt},clip,width=0.492\linewidth]{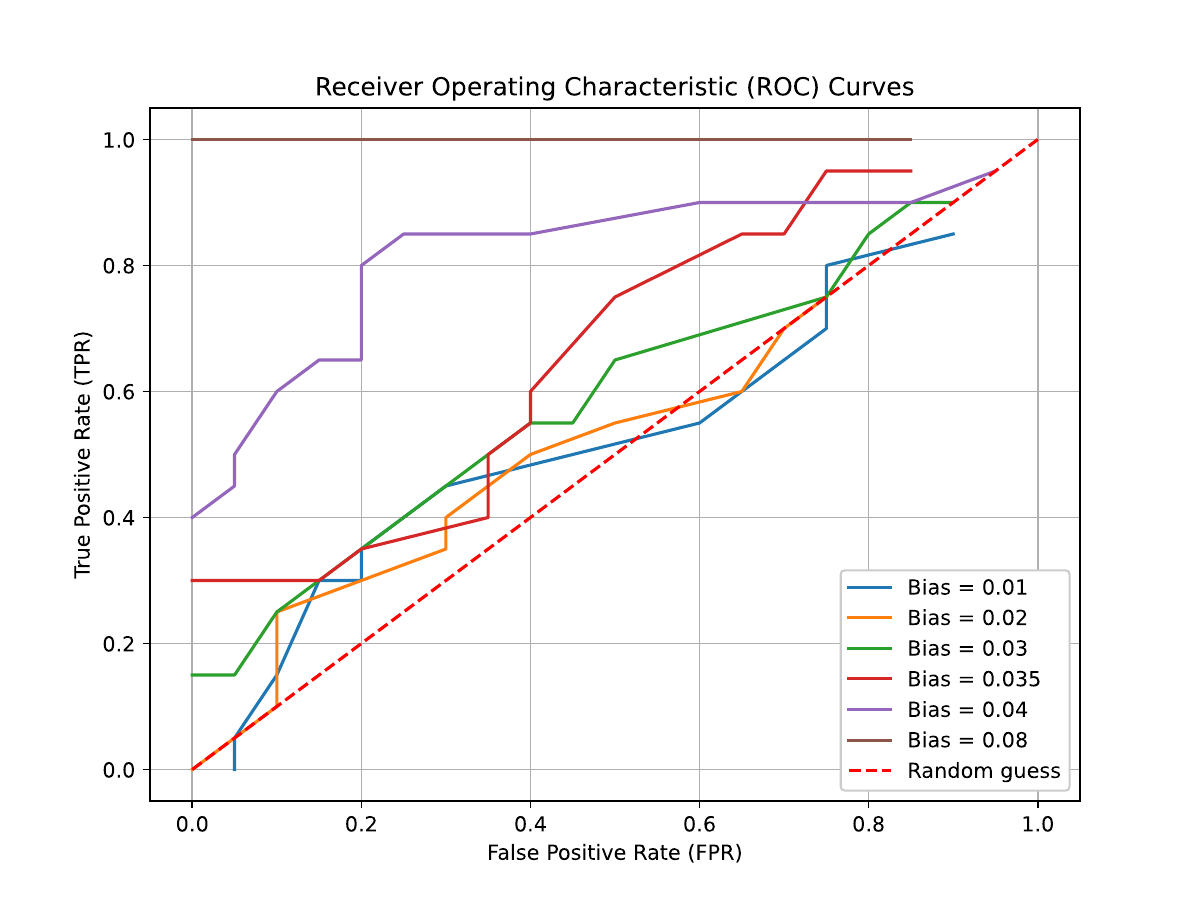} \\[0pt]
    \footnotesize{(a) Noise increase} \hspace{0.75in}
    \footnotesize{(b) Velocity-slip bias}
    \caption{ROC curves over fault-magnitude sweeps. As measurement noise and bias increase, discriminability improves and AUC approaches one, matching residual-test theory.}
    \label{fig:roc_overview}
\vspace{-3pt}
\end{figure}

Fig.~\ref{fig:roc_overview} summarizes discriminability over fault-magnitude sweeps.
For measurement-noise increases depicted in Fig.~\ref{fig:roc_overview}(a), the Receiver Operating Characteristic (ROC) curve approaches the upper-left corner as the noise standard deviation exceeds the practical threshold and yields near-unity Area Under the Curve (AUC).
For velocity-slip bias, shown in Fig.~\ref{fig:roc_overview}(b), small biases are harder to detect, but moderate-to-large biases yield high true-positive rates at low false-positive rates.
Together, Fig.~\ref{fig:roc_overview} and Tables~\ref{table:noise_summary}-\ref{table:bias_summary_acc} show that the detector achieves timely declarations while the integration layer maintains task completion with limited reassignment.\looseness=-1

\begin{table*}[t]
\centering
\setlength{\tabcolsep}{5pt}
\caption{Measurement-noise scenario: detection step and mission outcomes. NIS statistics are reported for healthy robots.}
\label{table:noise_summary}
\begin{tabular}{ccccccccc}
    \toprule
    \textbf{Robot} &
    \makecell{\textbf{Faulty?}} &
    \makecell{\textbf{Task} \\ \textbf{Before Fault}} &
    \makecell{\textbf{Task} \\ \textbf{After Fault}} &
    \makecell{\textbf{Fault} \\ \textbf{Init. Step}} &
    \makecell{\textbf{Fault} \\ \textbf{Det. Step}} &
    \makecell{\textbf{Task} \\ \textbf{Completed}} &
    \makecell{\textbf{NIS} \\ \textbf{Average}} &
    \makecell{\textbf{NIS} \\ \textbf{Std. Dev}} \\
    \midrule
    0 & No  & Task 1 & Task 1 & N/A & N/A & Yes & 0.960 & 0.249 \\
    1 & No  & Task 0 & Task 2 & N/A & N/A & Yes & 0.832 & 0.193 \\
    2 & Yes & Task 2 & N/A    & 25  & 36  & No  & N/A   & N/A   \\
    3 & No  & N/A    & Task 0 & N/A & N/A & Yes & 0.903 & 0.470 \\
    \bottomrule
\end{tabular}
\end{table*}

\begin{table*}[t]
\vspace{-5pt}
\centering
\setlength{\tabcolsep}{5pt}
\caption{Velocity-slip bias scenario: detection step and mission outcomes under reassignment.}
\label{table:bias_summary_acc}
\begin{tabular}{ccccccccc}
    \toprule
    \textbf{Robot} &
    \makecell{\textbf{Faulty?}} &
    \makecell{\textbf{Task} \\ \textbf{Before Fault}} &
    \makecell{\textbf{Task} \\ \textbf{After Fault}} &
    \makecell{\textbf{Fault} \\ \textbf{Init. Step}} &
    \makecell{\textbf{Fault} \\ \textbf{Det. Step}} &
    \makecell{\textbf{Task} \\ \textbf{Completed?}} &
    \makecell{\textbf{NIS} \\ \textbf{Average}} &
    \makecell{\textbf{NIS} \\ \textbf{Std. Dev}} \\
    \midrule
    0 & No  & Task 2 & Task 2 & N/A & N/A & Yes & 1.095 & 0.315 \\
    1 & Yes & Task 1 & N/A    & 20  & 28  & No  & N/A   & N/A   \\
    2 & No  & Task 0 & Task 0 & N/A & N/A & Yes & 0.934 & 0.301 \\
    3 & No  & N/A    & Task 1 & N/A & N/A & Yes & 0.972 & 0.252 \\
    \bottomrule
\end{tabular}
\end{table*}

Across these scenarios, healthy-stream NIS averages remain near one, consistent with standard Kalman residual properties \cite{thrun2005probabilistic,PraFar2025}. Communication dropouts are the most challenging due to delayed observability; in those runs, health masks and slack maintain continuity until measurements resume.

\subsection{Sensitivity and Scaling}
Tables~\ref{tab:Accuracy_noise_recap} and \ref{tab:Accuracy_bias_recap} report detection accuracy versus fault magnitude. For measurement noise, accuracy rises sharply around a sensor standard deviation of approximately $0.010$-$0.011$, which indicates a practical threshold; beyond this knee, accuracy is near $100\%$. For velocity bias, the knee occurs around $0.036$-$0.040$, after which detection becomes reliable. Table~\ref{tab:faults_summary} consolidates thresholds and latency trends across fault types and notes operational caveats, such as tuning stall gates to avoid flagging planned pauses.
Detector updates have constant computational complexity per robot, and the allocation remains a QP with health-dependent parameters. In our runs, measured per-step times were small relative to the control step, which confirms real-time feasibility and that health weighting/masks do not alter solver scaling.

Videos of the experiments presented in this section can be found in the accompanying video for the article, available at: \texttt{ \url{https://vimeo.com/sfarzan/acc2026}}

\begin{table}[t]
\vspace{-12pt}
\centering
\caption{Detection accuracy versus measurement-noise standard \\[-2pt] deviation. The knee occurs near $0.010$-$0.011$.}
\label{tab:Accuracy_noise_recap}
\begin{tabular}{cc||cc}
\toprule
\textbf{Noise Level} & \textbf{Accuracy (\%)} & \textbf{Noise Level} & \textbf{Accuracy (\%)} \\
\midrule
0.0070 & 6.67  & 0.0102 & 93.3 \\
0.0074 & 20.0  & 0.0106 & 96.6 \\
0.0078 & 23.3  & 0.0110 & 100 \\
0.0082 & 23.3  & 0.0114 & 96.6 \\
0.0086 & 36.6  & 0.0118 & 100 \\
0.0090 & 46.6  & 0.0122 & 100 \\
0.0094 & 46.6  & 0.0126 & 100 \\
0.0098 & 80.0  & 0.0130$+$ & 100 \\
\bottomrule
\end{tabular}
\end{table}

\begin{table}[t]
\vspace{-10pt}
\centering
\caption{Detection accuracy versus velocity-slip bias. Reliability \\[-2pt] improves markedly beyond $0.036$-$0.040$.}
\label{tab:Accuracy_bias_recap}
\begin{tabular}{cc||cc}
\toprule
\textbf{Bias Level} & \textbf{Accuracy (\%)} & \textbf{Bias Level} & \textbf{Accuracy (\%)} \\
\midrule
0.000 & 8.0  & 0.024 & 20.0 \\
0.004 & 2.0  & 0.028 & 26.0 \\
0.008 & 0.0  & 0.032 & 42.0 \\
0.012 & 12.0 & 0.036 & 76.0 \\
0.016 & 6.0  & 0.040 & 96.0 \\
0.020 & 12.0 & 0.044$+$ & 100 \\
\bottomrule
\end{tabular}
\end{table}

\begin{table}[t]
\vspace{-12pt}
\centering
\renewcommand{\arraystretch}{0.9}
\footnotesize
\caption{Fault-type summary: thresholds, detection-latency trends, \\[-2pt] and operational notes observed in experiments.}
\label{tab:faults_summary}
\begin{tabular}{|l|>{\raggedright\arraybackslash}p{5.8cm}|}
\hline
\textbf{Fault Category} & \textbf{Performance Characteristics and Limitations} \\
\hline
\makecell[l]{Measurement\\Noise Increase} &
\textit{Threshold:} $\sim$34\% above nominal standard deviation; \textit{Delay:} rapid once threshold is crossed; \textit{Note:} subtle increases can be sub-threshold. \\
\hline
\makecell[l]{Complete Task\\Abandonment} &
\textit{Threshold:} $\sim$4 steps of immobilization; \textit{Delay:} very fast (within 4 steps); \textit{Note:} tune to avoid flagging short planned pauses. \\
\hline
\makecell[l]{Measurement\\Bias} &
\textit{Threshold:} $\sim$3.6\% systematic error; \textit{Delay:} moderate (8-15 steps depending on magnitude); \textit{Note:} adaptive $Q$ can initially mask small biases. \\
\hline
\makecell[l]{Communication\\Failure} &
\textit{Mechanism:} covariance-growth monitoring; \textit{Delay:} long (tens of steps); \textit{Note:} no measurement validation during outage. \\
\hline
\makecell[l]{Stale Progress} &
\textit{Parameters:} $|\dot r|<\beta$ for $L$ steps; \textit{Behavior:} separates energy-aware pauses when $\beta,L$ are tuned; \textit{Note:} sensitivity depends on $\beta,L$. \\
\hline
\end{tabular}
\end{table}

\section{Conclusions and Future Directions}
We presented a task-progress monitor and a health-aware allocation scheme as a progress-based fault detection module for heterogeneous multi-robot teams. The detector models task progress with a lightweight KF and a NIS test augmented by rate and uncertainty gates and debounce. The main findings are threefold. First, the windowed NIS detector, augmented with stall and uncertainty gates, detects measurement-noise and velocity-bias faults with practical thresholds and delays consistent with its statistical design. Second, using health-weighted costs together with hard masks and an $\ell_1$ reassignment penalty preserves task coverage better than immediate exclusion alone while keeping reassignment limited. Third, communication dropouts are governed by covariance growth: the framework remains conservative under prolonged outages by suspending informative declarations once uncertainty becomes too large.

Future work includes decentralized integration where detection and allocation are performed on-board with limited communication, robustness under simultaneous multi-robot faults, and hardware validation in long-horizon field trials.
Moreover, the proposed monitor operates at the task-progress level rather than the physics level. This choice makes the detector portable across heterogeneous robots and across tasks for which a common state-space model is not available. It should therefore be viewed as complementary to platform-specific fault detection and isolation (FDI) modules: a low-level detector can identify component faults, whereas the proposed monitor measures whether those faults materially reduce task completion.
Finally, tasks that require multiple robots can be handled by moving from per-robot progress signals to shared task-level progress variables and by replacing one-robot assignment limits with team-size or coupling constraints. Designing such coupled detectors and allocators is an important direction for future work.

\bibliographystyle{IEEEtran}
\bibliography{refs}

\end{document}